\documentclass[fleqn,twoside]{article}
\usepackage{espcrc2}
\usepackage{amssymb}

\usepackage{graphicx}
\usepackage{epsfig}
\def\simle{%  ``less than about'' symbol
    \mathrel{\rlap{\raise 0.511ex
        \hbox{$<$}}{\lower 0.511ex \hbox{$\sim$}}}}
\def\lsim{\mathrel{\rlap{\lower4pt\hbox{\hskip1pt$\sim$}}
    \raise1pt\hbox{$<$}}}                % less than or approx. symbol
\def\gsim{\mathrel{\rlap{\lower4pt\hbox{\hskip1pt$\sim$}}
    \raise1pt\hbox{$>$}}}                % greater t
\newcommand{\vev}[1]{\left\langle#1\right\rangle}

\newcommand{\AmS}{{\protect\the\textfont2
  A\kern-.1667em\lower.5ex\hbox{M}\kern-.125emS}}

\newcommand{\be}{\begin{equation}}
\newcommand{\ee}{\end{equation}}
\newcommand{\beqn}{\begin{eqnarray}}
\newcommand{\eeqn}{\end{eqnarray}}

\title{
Hadron Spectrum and Decay Constant from $N_F=2$ Domain Wall QCD
}

\author{Taku Izubuchi\address{RIKEN-BNL Research Center, 
Brookhaven National Laboratory, Upton, New York 11973, USA}
\hspace*{-1.5mm}${}^{,}$\hspace*{-1mm}
\address{Institute for Theoretical Physics,
Kanazawa University, Kanazawa 920-1192, Japan} 
[RBC Collaboration]
\thanks{
We thank RIKEN, Brookhaven National Laboratory and the U.S. Department of 
Energy for providing the facilities essential for the completion of this work. 
This work is supported by RIKEN Super Combined Cluster at RIKEN. 
}
}
\begin{document}

\begin{abstract}
We report on the first large-scale study of two flavor QCD
with domain wall fermions (DWF). 
Simulation has been carried out at three dynamical quark mass
values about 1/2, 3/4, and 1 $m_{strange}$ on $16^3\times 32$
volume with $L_s=12$ and $a^{-1}\approx 1.7$ GeV. 
After discussing the details of the simulation, we report on the
light hadron spectrum and decay constants. 
\end{abstract}

\maketitle

\section{Introduction}
% To evaluate the systematic error in the quenched approximation of QCD, 
We have continued to investigate two flavor QCD using dynamical DWF. 
We present results of the hadron spectrum and the decay constants
as a project of the RBC collaboration.
For other quantities on the same dynamical ensembles, 
see other RBC contributions in these proceedings. 

%In practice, dynamical  simulation using domain wall fermion
%demands large comuputational resources  for its five dimensional extent, 
%$L_s$, and thus $L_s \sim O(10)$ is preferable. 
%On the other hand  explicit chiral symmetry breaking, 
%measured by the residual mass, $m_{res}$, for example, 
%must be small to realize the advantages of DWF. 

In this work we choose the DBW2 gauge action at $\beta=0.80$, 
in which the negative coefficient to the rectangular plaquette
suppresses dislocations of configuration and drastically 
reduces the residual mass, $m_{res}$, for relatively small 
$L_s \sim O(10)$ in quenched simulation\cite{Aoki:2002vt}. 
We simulated three sea quark mass $m_{sea}=0.02, 0.03$, and 0.04
using $N_F=2$ domain wall fermion with fifth dimensional extent $L_s=12$
and domain wall height $M_5=1.8$. 
%$L_s$,
$L_s=12$
%To keep the  scale of low energy physics approximately same, 
%$\beta$ is decreased to 0.80 from quenched simulation,
%$\beta=1.04$ of similar scale, and that leads to  rougher 
%gauge configuration at short distance. 
% as checked by measuring 
%the effective gauge action for  small size observables by 
%solving the Schwinger-Dyson equation in \cite{DynPaper}. 

%%   Among other observations, one of the worst systematic errors in quenched
%% simulations is revealed in the non-singlet flavor scalar meson correlator, 
%% $a_0$ meson, using Wilson fermions \cite{Bardeen:2003qz}, and
%% partially quenched DWF \cite{Prelovsek:2004jp}. 
%% The connected point-to-point scalar correlator, $C_{a_0}(t)$, 
%% was found to be negative for small $t$ when valence quark mass,
%% $m_{val}$, is smaller than sea quark mass, $m_{sea}$, 
%% which is a clear signal of the breakdown of the unitarity due to 
%% the quenching. 
%% %In the partially quenched simulation, the negative $C_{a_0}(t)$ at small 
%% %$t$ becomes positive when valence quark mass, $m_{val}$, 
%% %is larger than sea quark mass, $m_{sea}$. 

%% As an attept to coupe with such systematic errors due to the quenching
%% as well as the other unphysical symmetry breakings, 
%% dynamical DWF is  one of the natural steps to make. 
%\vspace*{-0.45cm}
\section{Ensemble generation}

\begin{table*}%[h]
\caption{ The details of $L_s=12, M_5=1.8$ DWF HMC-$\Phi$ evolution 
on $16^3\times 32$ lattice with $\beta=0.80$ DBW2 action. 
}
\begin{center}
\begin{tabular}{lcccccccccccc}\hline \hline
 $m_{\rm sea}$    &  $\Delta t$ & Steps/Traj.& Traj. & $P_{acc}$ &
$C_{\Delta H}$ & $N^{(0)}_{CG}$ & $N^{(7)}_{CG}$ & $N^{tot}_{CG}$ & 
$\tau_{int}$ \\ 
\hline
$0.02$   &  $1/100$    &  51          &  5361      & 77\% &
16.2(2) &  715 &  277 & 16014   &  2.9(8) \\
\hline
 $0.03$   &  $1/100$   &  51          &  6195      & 78\% &
15.8(1) &  514  & 158   &  9214 & 3.3(5) \\
\hline
 $0.04$   &  $1/80$    &  41           &  5605        & 68\% &
16.4(2) &  402  &  121  &  5964 & 4.5(10) \\ 
\hline \hline
\end{tabular}
\end{center}
\label{tab:evols}
\vspace*{-0.5cm}
\end{table*}
Table~\ref{tab:evols} summarizes the HMC-$\Phi$ evolution. 
An interesting result  is that
the acceptance, $P_{acc}$,  is insensitive to $m_{sea}$. 
More precisely, at least for our particular run parameters: 
$m_{sea} \gsim m_{strange}/2$, the squared energy difference 
between the first and the last configuration in a trajectory, 
scaled  by $V$ and the MD step size, $\Delta t$, 
\be
C_{\Delta H} = \sqrt{ \vev{(\Delta H)^2}\over V  (\Delta t)^4 } ~~, 
%P_{acc} \approx {\rm erfc}(\sqrt{\vev{(\Delta H)^2}/8})
\ee
stays same for all sea quark masses in the simulation, 
while an empirical estimation for staggered or Wilson fermions 
would be $C_{\Delta H} \sim m_{sea}^{-\alpha}, \alpha\sim 2$, 
in which case number of steps has to be increased 
proportional to $m_{sea}^{-1}$ to keep the acceptance 
$P_{acc} \approx {\rm erfc}(\sqrt{\vev{(\Delta H)^2}/8})$ constant. 
By introducing the new force term described in \cite{Dawson:2003ph}, 
$C_{\Delta H}$ is reduced to $\sim$ 40 \% of the old one.

We have implemented the chronological inverter technique 
of~\cite{Brower:1997vx}, in which the starting vector of the
conjugate gradient algorithm (CG) in a MD step is forecasted  
by the solutions in the previous steps. 
$N^{(0)}_{CG}$ in the table is the number of matrix multiplication 
in CG without forecasting, 
while $N^{(7)}_{CG}$ is the count using the previous seven 
solutions. The number seven is about the point where the precision of 
the forecast is saturated. 
$N^{({tot})}_{CG}$ is the total number of CG count in a trajectory. 
By a scaling ansatz, $N^{(i)}_{CG} = C_i (m_{sea}+m_{res})^{-\beta_i}$, 
$\beta_0 \sim 1$ and $\beta_7 \sim 1.5$ from the table. 
Although $\beta_7 > \beta_0$, the coefficient, $C_i$, is much smaller
for forecasting and thus $N^{(tot)}_{CG}$ is reduced roughly by  factors 
of two to three in the simulation points compared to the case without
the forecast. 

$\tau_{int}$ in the table is an estimation for the integrated 
auto-correlation length for $1\times1$ plaquette 
using whole trajectories by the truncated sum at 50 
with an error from jackknife blocks of length 100. 
$\tau_{int}$ are all equal within the quoted errors. 
The autocorrelation length of axial current correlator measured 
every ten trajectory is $\lsim 50$ trajectories\footnote{ 
The topological charge and the chiral condensation 
(with many random hits) show longer autocorrelation times.} 

Although these observations are encouraging for
DWF simulations with lighter quark masses in the future, 
we note that they may only be true for relatively heavy $m_{sea}$.

%%%%%%%%%%%%%%%%%%%%%%%%%%%%%%%%%%%%%%%
\section{Physical Results}
For each of three $m_{sea}$, we measure on  94 lattices separated 
by 50 trajectories, leaving out the first $\simeq$ 600 trajectories 
to allow the evolution to thermalized. 
All quoted errors are from jackknife estimate of 
the statistical error, correlated fit to hadron propagators 
is made using a single covariance matrix computed on the
entire ensemble. 

By a linear diagonal extrapolation, $m_{sea}=m_{val}\to 0$, 
we found $m_{res}=0.00137(4)$ or $\lsim$ 5 MeV, which is larger
than quenched DBW2 values for the same $L_s$ and $a^{-1}$, 
but it is still an order of magnitude smaller than 
the input quark  mass. 
The impacts of $m_{res}$ on the operator mixing for $B_K$ is
discussed elsewhere\cite{ChrisLattice2004,DynPaper}. 

From non-gauge-fixed wall-point pseudo-scalar correlator, 
$\vev{J^a_5 J^a_5}$ we extract  pseudo-scalar decay constant: 
\be
f_{ps} = 
{ 2 \over m_{ps}^2 }
(m_f+m_{res}) \vev{0 | J^a_5 |\pi,\vec{p}=0}~~,
\ee
and linearly extrapolate to the chiral limit, 
\def\mqval{m^q_{val}}
\def\mqsea{m^q_{sea}}
\def\mqvalsea{m^q_{val,sea}}
$\mqval = \mqsea = 0$ with 
$\mqval \equiv m_{sea}+m_{res},~ \mqsea=m_{val}+m_{res}$, 
to obtain an estimation for its chiral value, $f=0.078(1)$. 
The fit to the next-to-leading order (NLO) 
partially quenched chiral perturbation theory (PQChPT)
formula\cite{Golterman:1997st,Laiho:2002jq} for $f_{ps}$ 
%didn't reconstruct the data neighbor on the fit range very well.
%does not fit the data too well\cite{DynPaper}. 
did not describe the data in the neighborhood of
the fit range too well\cite{DynPaper}. 
% reconstruct the data neighbor on the fit range very well.
However, the following results depend on the value of $f$ only
mildly, so we will use the estimation from the linear fit.

Fitting Coulomb gauge-fixed wall-point correlators, we obtain
the mass of the pseudo-scalar mesons and the vector mesons. 
It is worth mentioning that a simple linear extrapolation 
of $m_{ps}^2$ to $\mqvalsea = 0$ is zero within the statistical error. 
This did not happen in the quenched case:  $m_{ps}^2= 0$
at $m_f\approx -(2-3)\times m_{res}$\cite{Aoki:2002vt}. 
This is consistent  with the difference between 
chiral logarithms ($m_{ps}^2/m^q \sim 2 B_0  + c m^q \log m^q$) vs 
the quenched one ($m_{ps}^2/m^q \sim \log m^q$) at small $m^q$. 

\begin{figure}[h]
\begin{center}
\includegraphics[scale=0.3]{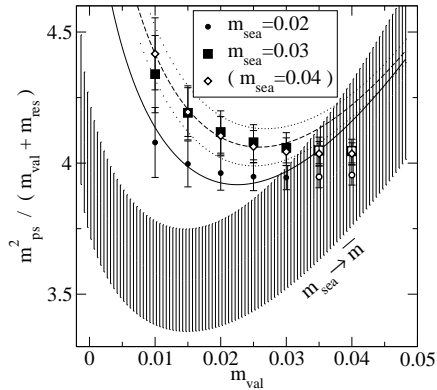}
\end{center}
\vspace{-18mm}
\caption{$m^2_{ps}/(m_{val}+m_{res})$ as a function of 
$m_{val}$ for each $m_{sea}=0.04, 0.03, 0.02$.
The curves are from the results of NLO fit 
using $m_{sea,val}\le 0.03$.
}
\label{fig:nlo_extr}
\vspace*{-4mm}
\end{figure}

The NLO PQChPT formula, 
{\arraycolsep=0.2pt
\begin{eqnarray}
&&M^{2}_{\pi,K (1-loop)}=M^{2}\left(1+\frac{\Delta M^2}{M^2}\right)
\label{eq:nlo mpisq}\\
%f_\pi&=&f_K=f\,\left(1+\frac{\Delta f}{f}\right)\\
%\label{eq:nlo decay}
&&\frac{\Delta M^2}{M^2}  =
\frac{2}{N}\!\!\left[\frac{M^2\!-\!M^2_{SS}}{16\pi^2f^2}+\frac{2M^2-M^2_{SS}}{M^2}
A_0(M^2)\right] \nonumber\\
&&\ \ \ - \frac{16}{f^2} [(L_5 - 2L_8)M^2 %\nonumber \\ &&
+ (L_4-2L_6) NM^2_{SS}],
\label{eq:nlo fit 1}%\\
%\frac{\Delta f}{f} & = & -NA_0(M^2_{vS}) + \frac{8}{f^2} (L_5 M^2
%+ L_4 NM^2_{SS}),
%\label{eq:nlo fit 2}
\end{eqnarray}
}
with
\begin{eqnarray}
M^2 &=& {2\,B_0}\,\mqval~,~ M^2_{SS} = {2\,B_0}\,\mqsea~,\\
%M^2_{vS} &=& \frac{(M^2+M^2_{SS})}{2},\\
A_0(M^2)&= &\frac{1}{16\pi^2f^2} M^2 \ln \frac{M^2}{\Lambda_\chi^2}~~,
\end{eqnarray}
is used to fit the pseudo-scalar mass obtained from axial current correlators
and pseudo-scalar correlators for various mass ranges. 
The NLO fit is constrained as $m_{ps}^2=0$ at $m_{val}=-m_{res}$, 
as it must due to the universality of the low energy domain 
wall fermion theory, and $f$ estimated above is used as an input. 
The results of fit of the fit are shown 
in Figure~\ref{fig:nlo_extr} 
and in Table~\ref{tab:nlo mpisq fit}. 
Note that $B_0$ is stable under change of fitting range, while the
$L_i$ coefficients are not.  
When the heavier mass is included in the fit, the $\chi^2/dof$ 
increases. 
\begin{table}[htdp]
\caption{Parameters from chiral perturbation theory fits to the values of
$m_{ps}^2$ at $m_f =m_{sea,val} \le m_f^{(max)}$ 
computed from the pseudo-scalar wall-point (upper two column),
and axial-vector wall point. $\chi^2$ is from uncorrelated fits in $m_f$.
$L_i$ refer to Gasser-Leutwyler low energy constants multiplied by
$10^4$ at $\Lambda_\chi=1$ GeV.
  }
\begin{center}
\begin{tabular}{ccccccc}\hline \hline
$m_f^{(max)}$ &  $\chi^2/$dof &  $2\,B_0$ &  $L_5-2L_8$ &
$L_4-2L_6$ \\\hline
 0.03 & $0.1(1)$  & $4.0(3)$ & $-1.5(7)$ & $-2(1)$  \\
 0.04 &  $2(1)$   & $4.2(1)$ & $-0.2(4)$ & $-1.1(4)$  \\ 
\hline
 0.03 &  $0.3(2)$ & $4.0(3)$ & $-1.9(8)$ & $-1(1)$  \\
 0.04 & $1.9(9)$  & $4.2(1)$ & $-0.4(4)$ & $-0.8(3)$  \\
\hline \hline
%0.01-0.03 & 9-16 & 0.14 (20) && 3.95 (20) & -0.00019 (10) &  -0.00015 (7) \cr
%0.01-0.04 & 9-16 & 2.03 (102) && 4.18 (12) & -0.00011 (3) &  -0.00002 (4) \cr
%0.02-0.04 & 9-16 & 1.17 (228)& -0.005 ( 4) & 4.19 (11) \cr
%\hline
%0.01-0.03 & 9-16 & 0.26 (24) && 4.03 (22) & -0.00012 (10) &  -0.00019 (8) \cr
%0.01-0.04 & 9-16 & 1.95 (83) && 4.23 (12) & -0.00008 (3) &  -0.00004 (5) \cr
%0.02-0.04 & 9-16 & 0.59 (155)& -0.003 ( 3) & 4.13 (11) \cr
%\hline
%0.01-0.03 & 9-16 & 0.48 (53) & &4.25 (22) & -0.00020 (8) &  0.00008 (6) \cr
%0.01-0.04 & 9-16 & 4.04 (230) & 4.70 (13) & -0.00012 (2) &  0.00026 (3) \cr
%\hline
%0.01-0.04 & 9-16 & 1.90 (170) && 4.31 (25) & -0.00018 (4) &  0.00017 (9) \cr
\end{tabular}
\end{center}
\label{tab:nlo mpisq fit}
\vspace*{-1cm}
\end{table}%

A linear extrapolation of the three vector meson mass points 
$m_{val}=m_{sea}$, yields $a^{-1}=$1.69(5) GeV at the 
quark mass, $m_f=\bar m=0.0002(1)$ corresponding to the neutral 
pion mass using (\ref{eq:nlo mpisq}). Similarly, but
using the NLO ChPT prediction for non-degenerate valence quarks, 
$m_s=0.045(3)$ is obtained from the physical kaon mass. 
The coefficients in the NLO formula for non-degenerate 
quarks are determined by the fit of the degenerate quark meson. 
It is important to note that $\bar m$ and $m_s$ as defined 
above are bare quark masses; the renormalized 
quark mass is defined as $Z_m m^q = Z_m(m+m_{res})$, where
$Z_m$ is a scheme and scale dependent renormalization 
factor\cite{Dawson:2003ph}. 

Extrapolating the measured decay constant at the partially
quenched points ($m_{val,sea} \le 0.04$) using the linear ansatz
%\be
%  f_{ps} = f + c_1(m_{v1}+m_{v2})/2 + c_2 m_{sea}~~, 
%\ee
to  $\bar m$ and $m_s$, we find $f_\pi =134(4)$ MeV, 
$f_K=157$(4) MeV, and $f_K/f_\pi =1.18(1)$, which agree
better with experiment than quenched DWF simulations. 

Sommer scale, $r_0$ (with $r_0=0.5$fm ), 
from the static quark potential gives almost
the same value for the lattice spacing 
\cite{KoichiLattice2004}. 
A similar analysis has been carried out for baryons, and
the chiral limit of the linear extrapolation gives $M_N/M_\rho$ = 1.34(4),
which is larger than the experimental value. This would be expected
from the relatively small spatial volume. 
 
\section{Conclusion}
  We have generated configurations of lattice QCD with two-flavor
dynamical DWF at $a^{-1}\approx 1.7$ GeV, with a spacial 
volume $V\approx (1.9 {\rm fm})^3$. 
From the $\sim 3 \times 5000$ trajectories in this work obtained with light
dynamical quarks and small residual quark mass, $m_{strange}/2 \simle
m_{sea}+m_{res} \simle m_{strange}$ or $ 0.54 \simle m_{ps}/m_V \simle 0.65$.  
We have tried the NLO ChPT fit for the mass and a simple linear fit for
the decay constant. The results for the physical decay constants are
closer than those in quenched DWF simulation. 
For further details of the simulations and results, we refer the reader 
to the forthcoming paper\cite{DynPaper}. 

\vspace*{-0.3cm}
\bibliography{bibfile}

\begin{thebibliography}{1}

\bibitem{Aoki:2002vt}
Y. Aoki et~al.,
\newblock Phys. Rev. D69 (2004) 074504, hep-lat/0211023.

\bibitem{Dawson:2003ph}
RBC, C. Dawson,
\newblock Nucl. Phys. Proc. Suppl. 128 (2004) 54, hep-lat/0310055.

\bibitem{Brower:1997vx}
R.C. Brower et~al.,
\newblock Nucl. Phys. B484 (1997) 353, hep-lat/9509012.

\bibitem{ChrisLattice2004}
RBC, C. Dawson,
\newblock in these proceedings.

\bibitem{DynPaper}
RBC, Y. Aoki et~al.,
\newblock forthcoming.

\bibitem{Golterman:1997st}
M.F.L. Golterman and K.C. Leung,
\newblock Phys. Rev. D57 (1998) 5703, hep-lat/9711033.

\bibitem{Laiho:2002jq}
J. Laiho and A. Soni,
\newblock Phys. Rev. D65 (2002) 114020, hep-ph/0203106.

\bibitem{KoichiLattice2004}
RBC, K. Hashimoto and T. Izubuchi,
\newblock in these proceedings.

\end{thebibliography}
\bibliographystyle{h-elsevier}
\end{document}